\documentclass[12pt]{article}

\begin{document}

\begin{center}
{\large{\bf Fuzzy Non-Trivial Gauge Configurations }}

\bigskip 
Badis Ydri \\
{\it Physics Department, Syracuse University, \\
Syracuse,N.Y.,13244-1130, U.S.A.}

\end{center}
\begin{abstract}
In this talk we will report on few results of discrete physics on the fuzzy sphere . In particular
non-trivial field configurations such as monopoles and solitons are constructed on fuzzy ${\bf S}^2$ using the
language of K-theory , i.e projectors . As we will show , these configurations are intrinsically finite
dimensional matrix models . The corresponding monopole charges and soliton winding numbers are also found using
the formalism of noncommutative geometry and cyclic cohomology .
\end{abstract}

Fuzzy physics is aimed to be an alternative method to approach discrete physics . Problems of lattice physics especially those with topological roots are all avoided on fuzzy spaces . For example , chiral anomaly , Fermion doubling and the discretization of non-trivial topological field configurations were all formulated consistently on the fuzzy sphere [see \cite{mythesis} and the extensive list of references therein] . The paradigm of fuzzy physics is ``discretization by quantization``, namely given a space , we treat it as a phase space and then quantize it . This requires the existence of a symplectic structure on this space . One such class of spaces which admit symplectic forms are the co-adjoint orbits, for example ${\bf C}{\bf P}^1={\bf S}^2$ , ${\bf C}{\bf P}^2$ , ${\bf  CP}^3$ and so on . Their quantization to obtain their fuzzy counterparts is done explicitly in \cite{balbook,mythesis} . Here we will only summarize the important results for ${\bf S}^2$ which are needed for the purpose of this paper .

\section{Fuzzy ${\bf S}^2$}
Fuzzy ${\bf S}^2$ or ${\bf S}^2_F$ is the algebra ${\bf A}=Mat_{2l+1}$ of $(2l+1){\times}(2l+1)$ matrices which is generated by the operators $n_i^F$ , $i=1,2,3$ , which are defined by
\begin{equation}
n_i^F=\frac{L_i}{\sqrt{l(l+1)}}.
\end{equation}
$L_i$'s satisfy $[L_i,L_j]=i{\epsilon}_{ijk}L_k$ and $\sum_{i=1}^3L_i^2=l(l+1)$ respectively, where $l$ is a positive integer . In other words , $L_i$'s are the generators of the IRR $l$ of $SU(2)$ . A general element $\hat{f}$ of ${\bf A}$ admits an expansion , in terms of $n_i^F$'s , of the form $\hat{f}(\vec{n}^F)=\sum_{i_1,...,i_k}f_{i_1,...,i_k}n_{i_1}^F....n_{i_k}^F$, which will terminate by the nature of the operators ${n}_i^F$'s . The continuum limit is defined by $l{\longrightarrow}{\infty}$ . In such a limit the fuzzy coordinates ${n}^F_{i}$'s tend , by definition , to the commutative coordinates $n_i$'s [ by inspection the commutators of the fuzzy coordinates among each others vanish at $l{\longrightarrow}{\infty}$ , but from the Casimir equation above we must have $\sum_{i=1}^{3}n_i^2=1$ ] . Furthermore , the noncommutative algebra at this limit becomes the commutative algebra of functions on continuum ${\bf S}^2$ , namely ${\bf A}{\longrightarrow}{\cal A}$ , where a general element $f$ of ${\cal A}$ will admit the expansion ${f}(\vec{n})=\sum_{i_1,...,i_k}f_{i_1,...,i_k}n_{i_1}....n_{i_k}$ .

Viewing ${\bf S}^2$ as a submanifold of ${\bf R}^3$ , one can check the following basic identity\cite{gkp}
\begin{equation}
{\cal D}_2={\cal D}_3|_{r={\rho}}+\frac{i{\gamma}^3}{\rho}.\label{25}
\end{equation}
${\gamma}^{a}={\sigma}_a$, $a=1,2,3$ , are
the flat gamma matrices in $3-$dimensions . ${\cal D}_2$ , ${\cal D}_3$ are the Dirac operators on ${\bf S}^2$ and
${\bf R}^3$ respectively . ${\cal D}_3|_{r={\rho}}$ is the restriction of the Dirac operator on ${\bf R}^3$ to the
sphere $r={\rho}$ , where ${\rho}$ is the radius of the sphere , namely $\sum_{{a}=1}^3x^2_{a}={\rho}^2$ for any
$\vec{x}{\in}{\bf S}^2$ . The Clifford algebra on ${\bf S}^2$ is two dimensional and therefore at each point
$\vec{n}=\vec{x}/{\rho}$ one has only two independents gamma matrices , they can be taken to be ${\gamma}^1$ and
${\gamma}^2$ . ${\gamma}^3$ should then be identified with the chirality operator ${\gamma}=\vec{\sigma}.\vec{n}$
on ${\bf S}^2$ .

Next by using the canonical Dirac operator ${\cal D}_3=-i{\sigma}_a{\partial}_a$ in (\ref{25}) one can derive the two
following equivalent expressions for the Dirac operator ${\cal D}_2$ on ${\bf S}_2$ :
\begin{eqnarray}
{\cal D}_{2g}&=&\frac{1}{\rho}(\vec{\sigma}\vec{\cal L}+1)\nonumber\\
{\cal D}_{2w}&=&-\frac{1}{\rho}{\epsilon}_{ijk}{\sigma}_in_j{\cal J}_k.\label{26}
\end{eqnarray}
${\cal L}_k=-i{\epsilon}_{kij}x_i{\partial}_j$ is the orbital angular
momentum and ${\cal J}_k={\cal L}_k+\frac{{\sigma}_k}{2}$ is the total angular momentum . $g$ and $w$ in (\ref{26})
stands for Grosse-Klim\v{c}\'{i}k-Pre\v{s}najder \cite{gkp} and Watamuras Dirac operators \cite{w} respectively . It is not difficult
to check that ${\cal D}_{2w}=i{\gamma}{\cal D}_{2g}={\cal D}_3|_{r=\rho}+\frac{i{\gamma}}{\rho}$ which means that
${\cal D}_{2w}$ and ${\cal D}_{2g}$ are related by a unitary transformation and therefore are equivalent. The
spectrum of these Dirac operators is trivially derived to be given by ${\pm}\frac{1}{\rho}(j+\frac{1}{2})$ where
$j$ is the eigenvalue of $\vec{\cal J}$ , i.e $\vec{\cal J}^2=j(j+1)$ and $j=1/2,3/2,...$ .

The fuzzy
versions of the Dirac operators (\ref{26}) are taken to be
\begin{eqnarray}
D_{2g}&=&\frac{1}{\rho}(\vec{\sigma}.ad\vec{L}+1)\nonumber\\
D_{2w}&=&\frac{1}{\rho}{\epsilon}_{ijk}{\sigma}_in_j^FL_k^R.\label{27}
\end{eqnarray}
$ad\vec{L}=\vec{L}^L-\vec{L}^R$ is the fuzzy derivation which annihilates the identity matrix in ${\bf A}$ as the
classical derivation $\vec{\cal L}$ annihilates the constant function in ${\cal A}$ . $\vec{L}^L$ and $-\vec{L}^R$ are the
generators of the IRR $l$ of $SU(2)$ which act on the left and on the right of the algebra ${\bf A}$ respectively , i.e $\vec{L}^Lf=\vec{L}f$ and $-L^R_if=-fL_i$ for any
$f{\in}{\bf A}$ . From this definition one can see that $Ad{L}_i$ provide the generators of the adjoint action of
$SU(2)$ on ${\bf A}$ , namely $Ad\vec{L}(f)=[\vec{L},f]$ for any $f{\in}{\bf A}$ .

These two fuzzy Dirac operators are not unitarily equivalent anymore . This can be checked by computing their
spectra . The spectrum of $D_{2g}$ is exactly that of the continuum only cut-off at the top total angular momentum
$j=2l+\frac{1}{2}$ . In other words the spectrum of $D_{2g}$ is equal to $\{{\pm}\frac{1}{\rho}(j+\frac{1}{2})$ ,
$j=\frac{1}{2},\frac{3}{2},...2l-\frac{1}{2}\}$ and $D_{2g}(j)=\frac{1}{\rho}(j+\frac{1}{2})$ for
$j=2l+\frac{1}{2}$ . The spectrum of $D_{2w}$ is , however , highly deformed as compared to the continuum spectrum
especially for large values of $j$ . It is given by
$D_{2w}(j)={\pm}\frac{1}{\rho}(j+\frac{1}{2})\sqrt{1+\frac{1-(j+1/2)^2}{4l(l+1)}}$ . From these results it is
obvious that $D_{2g}$ is superior to $D_{2w}$ as an approximation to the continuum .

In the same way one can find the fuzzy chirality operator ${\Gamma}$ by the simple replacement
$\vec{n}{\longrightarrow}\vec{n}^F$ in ${\gamma}=\vec{\sigma}.\vec{n}$ and insisting on the result to have the
following properties  : $1){\Gamma}^2=1$, ${\Gamma}^{+}={\Gamma}$ and $[{\Gamma},f]=0$ for all $f{\in}{\bf A}$ .
One then finds\cite{w}
\begin{equation}
{\Gamma}=\frac{1}{l+\frac{1}{2}}(-\vec{\sigma}\vec{L}^R+\frac{1}{2}).\label{28}
\end{equation}
Interestingly enough this fuzzy chirality operator anticommutes
with ${D}_{2w}$  and not with ${D}_{2g}$ so ${D}_{2w}$ is a better approximation to the continuum than
$D_{2g}$ from this respect . This is also clear from the spectra above , in the spectrum of $D_{2g}$ the top
angular momentum is not paired to anything and therefore $D_{2g}$ does not admit a chirality operator .

\section{Fuzzy Non-Trivial Gauge Configurations} 
\subsection{Classical Monopoles}
Monopoles are one of the most fundamental non trivial
configurations in field theory. The wave functions of a particle of charge $q$ in the field of a monopole $p$ ,
which is at rest at $r=0$ , are known to be given by the expansion \cite{balbook}
\begin{equation}
{\psi}^{(N)}(r,g)=\sum_{j,m}c_{m}^{j}(r)<j,m|D^{(j)}(g)|j,-\frac{N}{2}>,\label{29}
\end{equation}
where $D^{(j)}:g{\longrightarrow}D^{(j)}(g)$ is the $j$ IRR of $g{\in}SU(2)$ . The integer $N$ is related to $q$
and $p$ by the Dirac quantization condition : $N=\frac{qp}{2{\pi}}$ . $r$ is the radial coordinate of the relative
position $\vec{x}$ of the system , the angular variables of $\vec{x}$ are defined through the element
$g{\in}SU(2)$ by $\vec{\tau}.\vec{n}=g{\tau}_3g^{-1}$ , $\vec{n}=\vec{x}/r$ . It is also a known result that the
precise mathematical structure underlying this physical system is that of a $U(1)$ principal fiber bundle
$SU(2){\longrightarrow}{\bf S}^2$. In other words for a fixed $r=\rho$ , the particle $q$ moves on a sphere ${\bf
S}^2$ and its wave functions (\ref{29}) are precisely elements of ${\bf {\cal S}}({\bf S}^2,SU(2))$ , namely sections
of a $U(1)$ bundle over ${\bf S}^2$ . They have the equivariance property
\begin{equation}
{\psi}^{(N)}(\rho,ge^{i{\theta}\frac{{\tau}_3}{2}})=e^{-i{\theta}\frac{N}{2}}{\psi}^{(N)}(\rho,g),\label{equivariance}
\end{equation}
i.e they are not really functions on ${\bf S}^2$ but rather functions on $SU(2)$ because they clearly depend on the
specific point on the $U(1)$ fiber  . In this paper , we will only consider the case $N={\pm}1$ . The case
$|N|{\neq}1$ being similar and is treated in great detail in \cite{mythesis,myfirst}.

An alternative description of monopoles can be given in terms of K-theory and projective modules . It is based on
the Serre-Swan's theorem \cite{connes,landi'sgood}which states that there is a complete equivalence between vector bundles
over a compact manifold ${\bf M}$ and projective modules over the algebra ${C}(\bf M)$ of smooth functions on
${\bf M}$ . Projective modules are constructed from ${C}(\bf M)^{n}={C}(\bf M){\otimes}{\bf C}^n$ where $n$ is some
integer by the application of a certain projector $p$ in ${\cal M}_{n}({C}(\bf M))$ , i.e the algebra of
$n{\times}n$ matrices with entries in ${C}(\bf M)$ .

In our case ${\bf M}={\bf S}^2$ and $C(\bf M)={\cal A}\equiv$ the algebra of smooth functions on ${\bf S}^2$ . For
a monopole system with winding number $N={\pm}1$ , the appropriate projective module will be constructed from
${\cal A}^2={\cal A}{\otimes}{\bf C}^2$ . It is ${\cal P}^{(\pm 1)}{\cal A}^2$ where ${\cal P}^{(\pm 1)}$ is the
projector
\begin{equation}
{\cal P}^{(\pm 1)}=\frac{1{\pm}\vec{\tau}.\vec{n}}{2}.\label{continuumprojector}
\end{equation}
It is clearly an element of ${\cal M}_{2}(\cal A)$ and satisfies ${\cal P}^{{(\pm 1)} 2}={\cal P}^{(\pm 1)}$ and
${\cal P}^{{(\pm 1)}+}={\cal P}^{(\pm 1)}$. ${\cal P}^{(\pm 1)}{\cal A}^2$ describes a monopole system with
$N={\pm}1$ as one can directly check by computing its Chern character as follows
\begin{equation}
\pm 1=\frac{1}{2{\pi}i}\int Tr{\cal P}^{({\pm}1)}d{\cal P}^{({\pm}1)}{\wedge}d{\cal P}^{({\pm}1)}.\label{winding1}
\end{equation}
On the contrary to the space of sections ${\cal S}({\bf S}^2,SU(2))$ , elements of ${\cal P}^{(\pm 1)}{\cal
A}^2$ are by construction invariant under the action $g{\longrightarrow}gexp(i{\theta}\frac{{\tau}_3}{2})$ . The
other advantage of ${\cal P}^{(\pm 1)}{\cal A}^2$ as compared to ${\cal S}({\bf S}^2,SU(2))$ is the fact
that its fuzzification is much more straight forward.

\subsection{On The Equivalence of ${\cal P}^{(\pm 1)}{\cal A}^2$ and ${\bf {\cal S}}({\bf S}^2,SU(2))$}
Before we start the fuzzification of ${\cal P}^{(\pm 1)}{\cal A}^2$ , let us first comment on the relation between
the wave functions ${\psi}^{(\pm 1)}$ given in equation (\ref{29}) and those belonging to ${\cal P}^{(\pm 1)}{\cal
A}^2$. The projector ${\cal P}^{(\pm 1)}$ can be rewritten as ${\cal P}^{(\pm 1)
}=D^{(\frac{1}{2})}\frac{1{\pm}{\tau}_3}{2}D^{(\frac{1}{2})+}(g)$ where
$D^{(\frac{1}{2})}:g{\longrightarrow}D^{(\frac{1}{2})}(g)=g$ is the $\frac{1}{2}$ IRR of $SU(2)$ . Hence ${\cal
P}^{(\pm 1)}D^{(\frac{1}{2})}(g)|\pm>=D^{(\frac{1}{2})}(g)\frac{1{\pm}{\tau}_3}{2}|\pm>=D^{(\frac{1}{2})}(g)|\pm>$
, where $|\pm>$ are defined by ${\tau}_3|\pm>=\pm|\pm>$ . In the same way one can show that ${\cal P}^{(\pm
1)}D^{(\frac{1}{2})}(g)|\mp>=0$ . This last result means that
\begin{eqnarray}
{\cal P}^{(\pm 1)}&=&D^{(\frac{1}{2})}(g)|\pm><\pm|D^{(\frac{1}{2})+}(g)\nonumber\\
\end{eqnarray}
$<\pm|D^{(\frac{1}{2})+}(g)$ defines a map from ${\cal P}^{(\pm 1)}{\cal A}^2$  into ${\cal S}({\bf
S}^2,SU(2))$ as follows
\begin{eqnarray}
<\pm|D^{(\frac{1}{2})+}(g)&:&|\psi>{\longrightarrow}<\pm|D^{(\frac{1}{2})+}(g)|\psi>={\psi}^{(\pm 1)}(\rho,g).
\end{eqnarray}
$<\pm|D^{(\frac{1}{2})+}(g)|\psi>$ has the correct transformation law (\ref{equivariance}) under
$g{\longrightarrow}gexp(i{\theta}\frac{{\tau}_3}{2})$ as one can check by using the basic equivariance property
\begin{equation}
D^{(\frac{1}{2})}(ge^{i{\theta}\frac{{\tau}_3}{2}})|\pm>=e^{\pm i \frac{\theta}{2}}D^{(\frac{1}{2})}(g)|\pm>.
\end{equation}
In the same way $D^{(\frac{1}{2})}(g)|\pm>$ defines a map , ${\cal S}({\bf S}^2,SU(2)){\longrightarrow}{\cal
P}^{(\pm 1)}{\cal A}^2$, which takes the wave functions ${\psi}^{(\pm 1)}$ to the two components elements
${\psi}^{(\pm 1)}D^{(\frac{1}{2})}(g)|\pm>$ of ${\cal P}^{(\pm 1)}{\cal A}^2$. Under
$g{\longrightarrow}gexp(i{\theta}\frac{{\tau}_3}{2})$ , the two phases coming from ${\psi}^{(\pm 1)}$ and
$D^{(\frac{1}{2})}(g)|\pm>$ cancel exactly so that their product is a function over ${\bf S}^2$ .
\subsection{Fuzzy Monopoles}
Towards fuzzification one rewrites the winding number (\ref{winding1}) in the form
\begin{eqnarray}
\pm 1 &=& -\frac{1}{4 \pi}\int d (\cos{\theta}){\wedge}d{\phi}\;{\rm Tr}\;{\gamma}{\cal P}^{(\pm 1)}\;[{\cal
D}, {\cal P}^{(\pm 1)}]\;[{\cal D}, {\cal P}^{(\pm 1)}](\vec{n})\nonumber\\
&=& -Tr_{\omega} \left( \frac{1}{|{\cal D}|^2}\gamma \;{\cal P}^{(\pm 1)}\; [{\cal D}, {\cal P}^{(\pm 1)}] \;[{\cal
D}, {\cal P}^{(\pm 1)}]\; \right).\label{winding2}
\end{eqnarray}
The first line is trivial to show starting from (\ref{winding1}) , whereas the second line is essentially Connes trace theorem \cite{connes} . $|{\cal D}|=$ positive square root of ${\cal D}^{\dagger}{\cal
D}$ while $Tr_{\omega}$ is the Dixmier trace \cite{connes,landi,varilly}.In the fuzzy setting , this
Dixmier trace will be replaced by the ordinary trace because the algebra of functions on fuzzy ${\bf S}^2_F$ is
finite dimensional .

${\cal D}$ in (\ref{winding2}) is either ${\cal D}_{2g}$ or ${\cal D}_{2w}$ which are given in equation (\ref{26}) . They
both give the same answer ${\pm}1$ . The fuzzy analogues of ${\cal D}_{2g}$ and ${\cal D}_{2w}$ are respectively
$D_{2g}$ and $D_{2w}$ given by equation (\ref{27}) . These latter operators were shown to be different and therefore
one has to decide which one should we take as our fuzzy Dirac operator . $D_{2g}$ does not admit as it stands a
chirality operator and therefore its use in the computation of winding numbers requires more care which is done in \cite{mythesis,mysecond} .$D_{2w}$ admits the fuzzy chirality operator (\ref{28}) which will be used
instead of the continuum chirality ${\gamma}=\vec{\sigma}.\vec{n}$. However $D_{2w}$ has a zero eigenvalue for
$j=2l+\frac{1}{2}$ so it must be regularized for its inverse in (\ref{winding2}) to make sense. This will be understood
but not done explicitly in this paper , a careful treatment is given in \cite{mythesis,myfirst}.

Finally the projector ${\cal P}^{(\pm 1)}$ will be replaced by a fuzzy projector $p^{(\pm 1)}$ which we will now
find  . We proceed like we did in finding the chirality operator ${\Gamma}$ , we replace $\vec{n}$ in (\ref{continuumprojector}) by
$\vec{n}^F={\vec{L}^L}/{\sqrt{l(l+1)}}$ and insist on the result to have the properties $p^{(\pm 1)2}=p^{(\pm 1)}$
and $p^{(\pm 1)+}=p^{(\pm 1)}$ . We also require this projector to commute with the chirality operator ${\Gamma}$,
the answer for winding number $N=+1$ turns out to be $p^{(+1)}=\frac{1}{2}+\frac{1}{2l+1}\big[\vec{\tau}.\vec{L}^L+\frac{1}{2}\big]$. This can be rewritten in the following useful form
\begin{equation}
p^{(+1)}=\frac{\vec{K}^{(1)2}-(l-\frac{1}{2})(l+\frac{1}{2})}{(l+\frac{1}{2})(l+\frac{3}{2})-(l-\frac{1}{2})(l+\frac{1}{2})},\label{fuzzyprojector1}
\end{equation}
where $\vec{K}^{(1)}=\vec{L}^L+\frac{{\vec{\tau}}}{2}$ . This allows us to see immediately that $p^{(+1)}$ is the
projector on the subspace with the maximum eigenvalue $l+\frac{1}{2}$ . Similarly , the projector $p^{(-1)}$ will
correspond to the subspace with minimum eigenvalue $l-\frac{l}{2}$ , namely
\begin{equation}
p^{(-1)}=\frac{\vec{K}^{(1)2}-(l+\frac{1}{2})(l+\frac{3}{2})}{(l-\frac{1}{2})(l+\frac{1}{2})-(l+\frac{1}{2})(l+\frac{3}{2})}.\label{fuzzyprojector2}
\end{equation}
By construction (\ref{fuzzyprojector1}) as well as (\ref{fuzzyprojector2}) have the correct continuum limit (\ref{continuumprojector}), and they are in the
algebra ${\cal M}_2(\bf A)$ where ${\bf A}$ is the fuzzy algebra on fuzzy ${\bf S}^2_F$, i.e
$2(2l+1){\times}2(2l+1)$ matrices . Fuzzy monopoles with winding number ${\pm}1$ are then described by the
projective modules $p^{(\pm 1)}{\bf A}^2$ .

If one include spin , then ${\bf A}^2$ should be enlarged to ${\bf A}^4$ . It is on this space that the Dirac
operator $D_{2w}$ as well as the chirality operator ${\Gamma}$ are acting . In the fuzzy the left and right actions
of the algebra ${\bf A}$ on ${\bf A}$ are not the same . The left action is generated by $L_i^L$ whereas the right action is
generated by $-L_i^R$  so that we are effectively working with the algebra ${\bf A}^L{\otimes}{\bf A}^R$ . A
representation ${\Pi}$ of this algebra is provided by ${\Pi}(\alpha)={\alpha}{\otimes}{\bf 1}_{2{\times}2}$ for any
${\alpha}{\in}{\bf A}^L{\otimes}{\bf A}^R$ . It acts on the Hilbert space ${\bf A}^4{\oplus}{\bf A}^4$ .

With all these considerations , one might as well think that one must naively replace
$Tr_{\omega}{\longrightarrow}Tr$ , ${\gamma}{\longrightarrow}{\Gamma}$ , ${\cal D}{\longrightarrow}D_{2w}$ and
${\cal P}^{(\pm 1)}{\longrightarrow}p^{(\pm 1)}$ in (\ref{winding2}) to get its fuzzy version . This is not totally correct
since the correct discrete version of (\ref{winding2}) turns out to be
\begin{equation}
c(\pm 1)=-Tr{\epsilon}P^{(\pm 1)}[F_{2w},P^{(\pm 1)}][F_{2w},P^{(\pm 1)}],\label{winding3}
\end{equation}
with
\begin{equation}
{\bf F}_{2w} = \left( \begin{array}{cc}
             0  & \frac{D_{2w}}{|D_{2w}|}\\
             \frac{D_{2w}}{|D_{2w}|}& 0
           \end{array} \right),
           \quad \epsilon=\left( \begin{array}{cc}
                                                    {\Gamma} & 0 \\
                                                        0  & {\Gamma}
                                                 \end{array} \right).\label{fredholm}
\end{equation}
and
\begin{equation}
P^{(\pm 1)}=\left( \begin{array}{cc}
           \frac{1+\Gamma}{2} p^{(\pm)} & 0 \\
                    0  & \frac{1-\Gamma}{2} p^{(\pm)}
         \end{array} \right). \label{projector}
\end{equation}
[For a complete proof see \cite{mythesis} or \cite{myfirst}].For $p^{(+1)}$ one finds that
$c(+1)=+1+[2(2l+1)+1]$ while for $p^{(-)}$ we find $c(-1)=-1+[2(2l)+1]$ . They are both wrong if compared to (\ref{winding2})!

The correct answer is obtained by recognizing that $c(\pm 1)$ is nothing but the index of the operator
\begin{equation}
\hat{f}^{(+)}=\frac{1-\Gamma}{2}p^{(\pm 1)}\frac{D_{2w}}{|D_{2w}|}p^{(\pm 1)}\frac{1+\Gamma}{2}.
\end{equation}
This index counts the number of zero modes of $\hat{f}^{(+)}$ . The proof starts by remarking that, by
construction , only the matrix elements $<p^{(\pm 1)}U_{-}|{\hat{f}}^{(+)}|p^{(\pm 1)}U_{+}>$ where
$U_{\pm}=\frac{1{\pm}\Gamma}{2}{\bf A}^4$ , exist and therefore ${\hat{f}}^{(+)}$ is a mapping from
$\hat{V}_{+}=p^{(\pm 1)}U_{+}$ to $\hat{V}_{-}=p^{(\pm 1 )}U_{-}$ . Hence
$Index{\hat{f}}^{(+)}=dim\hat{V}_{+}-dim\hat{V}_{-}$ .  

Since one can write the chirality operator ${\Gamma}$ in the form
${\Gamma}=\frac{1}{l+1/2}\Big[j(j+1)-(l+1/2)^2\Big]$ where $j$ is the eigenvalue of $(-\vec{L}^R +
\frac{\vec{\sigma}}{2})^2$ , $j=l\pm 1/2$ for which $\Gamma|_{j=l\pm 1/2}=\pm 1$ defines the subspace $U_{\pm}$
with dimension $2(l\pm 1/2)+1$ . On the other hand , for $p^{(+1)}$ which projects down to the subspace with
maximum eigenvalue $k_{max}=l+\frac{1}{2}$ of the operator $\vec{K}^{(1)}=\vec{L}+\frac{{\vec{\tau}}}{2}$ ,
$\hat{V}_{\pm}$ has dimension $[2(l{\pm}1/2)+1][2(l+1/2)+1]$ and so the index is
$Index{\hat{f}}^{(+)}=c(+1)=2(2l+2)$. This result signals the existence of zero modes of the operator ${\hat{f}}^{(+)}$ . Indeed for ${\Gamma}=+1$ one
must couple $l+\frac{1}{2}$ to $l+\frac{1}{2}$ and obtain $j=2l+1,2l,..0$, whereas for ${\Gamma}=-1$ we couple
$l+\frac{1}{2}$ to $l-\frac{1}{2}$ and obtain $j=2l,...,1$ . $j$ here denotes the total angular momentum
$\vec{J}=\vec{L}^L-\vec{L}^R+\frac{\vec{\sigma}}{2}+\frac{\vec{\tau}}{2}$ . Clearly the eigenvalues
$j^{(+1)}=2l+1$ and $0$ in $\hat{V}_+$ are not paired to anything . The extra piece in $c(+1)$ is therefore
exactly equal to the number of the top zero modes , namely $2j^{(+1)}+1=2(2l+1)+1$ . These modes do not exist in
the continuum and therefore they are of no physical relevance and must be projected out . This can be achieved by
replacing the projector $p^{(+1)}$ by a corrected projector ${\pi}^{(+1)}=p^{(+1)}[1-{\pi}^{(j^{(+1)})}]$ where
${\pi}^{(j^{(+1)})}$ projects out the top eigenvalue $j^{(+1)}$ , it can be easily written down explicitly . Putting ${\pi}^{(+1)}$ in (\ref{winding3}) gives exactly $c(+1)=+1$ which is the correct answer .

The same analysis goes for $p^{(-1)}$ . Indeed if we replace it by the  
corrected projector ${\pi}^{(-1)}=p^{(-1)}[1-{\pi}^{(j^{(-1)})}]$ where ${\pi}^{(j^{(-1)})}$ projects out the top
eigenvalue $j^{(-1)}=2l$, then equation (\ref{winding3}) will give exactly $c(-1)=-1$ which is what we want .
\section{Conclusion}
It was shown in this article that topological quantities can be precisely and strictly defined in the discrete setting by using the methods of noncommutative geometry and fuzzy physics .

\bibliographystyle{unsrt}

\begin{thebibliography}{10}

\bibitem{mythesis}
Badis Ydri , {\it{Fuzzy Physics}} , a thesis which will be submitted in partial fulfillment of the requirements for the degree of Ph.D in physics , syracuse university .
\bibitem{balbook}
A.P.Balachandran , G.Marmo ,B-S.Skagerstan and A.Stern, {\it{Classical Topology and Quantum States}}, World Scientific , Singapore , 1991 .

\bibitem{gkp}
H.~Grosse and P.~Pre\v{s}najder, {\em Lett.Math.Phys.} {\bf 33}, 171 (1995).
H.~Grosse, C.~Klim\v{c}\'{\i}k and P.~Pre\v{s}najder,{\em Commun.Math.Phys.} {\bf 178},507-526 (1996); {\bf 185}, 155 (1997);H.~Grosse and P.~Pre\v{s}najder,
{\em Lett.Math.Phys.} {\bf 46}, 61 (1998) and ESI preprint,1999. H.~Grosse, C.~Klim\v{c}\'{i}k, and P.~Pre\v{s}najder,{\em Comm.Math.Phys.} {\bf 180}, 429 (1996),{\tt hep-th/9602115}.H.~Grosse, C.~Klim\v{c}\'{i}k, and P.~Pre\v{s}najder,
in {\em Les Houches Summer School on Theoretical Physics}, 1995,{\tt hep-th/9603071} .


\bibitem{w}
U.~Carow-Watamura and S.~Watamura,{\tt hep-th/9605003},{\em Comm.Math.Phys.} {\bf 183}, 365 (1997), {\tt hep-th/9801195}.

\bibitem{myfirst}
S.~Baez, A.~P. Balachandran, S.~Vaidya and B.~Ydri,{\tt hep-th/9811169} and Comm.Math.Phys.{\bf 208},787(2000).

\bibitem{mysecond}
A.~P.~Balachandran, T.~R.~Govindarajan and B.~Ydri, {\em SU-4240-712,IMSc-99/10/36} and {\tt hep-th/9911087};A.~P.~Balachandran, T.~R.~Govindarajan and B.~Ydri, {\em SU-4240-712,IMSc-99/10/36} and {\tt hep-th/0006216} and {\em Mod.Phys.Lett.}{\bf A15}, 1279 (2000);

\bibitem{connes}
A.~Connes,{\em Noncommutative Geometry},Academic Press, London, 1994;


\bibitem{landi'sgood}
G.~Landi ."Deconstructing Monopoles and Instantons" . {\tt math-ph/9812004} .

\bibitem{landi}
G.~Landi,{\em An Introduction to Noncommutative spaces And Their Geometries},
Springer-Verlag, Berlin, 1997.{\tt hep-th/9701078}.

\bibitem{varilly}
J.~C. Varilly and J.~M. Gracia-Bondia.\newblock {\em J. Geom. Phys.}, 12:223--301, 1993.J.~C. Varilly.\newblock {\it An Introduction to Noncommutative Geometry}.\newblock {\em physics/9709045}.

\end{thebibliography}

\end{document}